\documentstyle[12pt]{article}
\makeatletter
\def\section{\@startsection {section}{1}{\z@}{3.5ex plus 1ex minus 
 .2ex}{2.3ex plus .2ex}{\large\bf}}
\def\lsim{\mathrel{\mathpalette\gl@align<}}
\def\gsim{\mathrel{\mathpalette\gl@align>}}
\def\gl@align#1#2{\lowe.6ex\vbox{\baselineskip\z@skip\lineskip\z@
    \ialign{$\m@th#1\hfil##\hfil$\crcr#2\crcr\sim\crcr}}}
\makeatother 

\newcommand{\newsection}[1]{
\setcounter{equation}{0}
\section{#1}
}
\newcommand{\be}{\begin{equation}}
\newcommand{\ee}{\end{equation}}
\newcommand{\bea}{\begin{eqnarray}}
\newcommand{\eea}{\end{eqnarray}}
\newcommand{\vs}[1]{\vspace{#1 cm}}

\def\f{\frac}

\def\d{\delta}

\def\l{\lambda}
\def\n{\nu}

\def\L{\Lambda}
\def\f#1#2{\frac{#1}{#2}}

\begin{document}
\begin{titlepage}
\vs{5}
\begin{center}
{\Large{\bf Analytic Solutions of The Wheeler-DeWitt Equation 
in Spherically Symmetric Space-time
       }
} 
\end{center}
\vspace{1cm}
\setlength{\footnotesep}{15pt}
\begin{center}
{\sc Masakatsu Kenmoku \footnote{e-mail:kenmoku@cc.nara-wu.ac.jp} 
, Hiroto Kubotani \footnote{e-mail:
kubotani@cpsun3.b6.kanagawa-u.ac.jp} 
, Eiichi Takasugi \footnote{e-mail:takasugi@phys.wani.osaka-u.ac.jp} 
 \\ and  Yuki Yamazaki \footnote{e-mail:yamazaki@phys.nara-wu.ac.jp} 
}\break
\\
{\it  $\ ^1$ Department of Physics, Nara Women's University,
~Nara 630,~Japan \\ 
 $\ ^2$ Department of Information Science, Kanagawa University,
~Yokohama 221,~Japan \\ 
 $\ ^3$ Department of Physics, Osaka University,~Osaka 560,~Japan \\
 $\ ^4$ Graduate School of Human Culture, Nara Women's University,
 ~Nara 630,~Japan  
}
\end{center}
\vs{1}
\begin{abstract}
We study the quantum theory of the Einstein-Maxwell action  
with a cosmological term in the spherically symmetric space-time, 
and explored quantum black hole solutions in 
Reissner-Nordstr$\ddot{{\rm o}}$m-de Sitter 
geometry. We succeeded to obtain analytic solutions 
to satisfy both the energy and momentum constraints.  
\end{abstract}
{\it keywords}:
Analytic solutions, Wheeler-DeWitt equations, 
quantum gravity, black hole 
\end{titlepage}
\newpage
%
\newsection{Introduction}
%

Combining the quantum theory and the general relativity is 
important 
for the unification of four elementary forces as well as for 
the smearing singularities appeared in the classical relativity 
and the exploring the fluctuation in the early universe. 
Black hole radiation is also one of important issues of 
quantum theory of gravity \cite{Hawking75}.  


Canonical formalism of general relativity was formulated by 
Arnowitt-Deser-Misner (ADM) \cite{ADM62} and by Dirac \cite{Dirac64}.  
According to the invariance of general coordinate transformation, 
Hamiltonian dynamics leads to a totally constrained system. 
A canonical Hamiltonian treatment has been developed for 
Schwarzschild geometry by Berger et al. \cite{Berger72} and 
Lund \cite{Lund73}, by taking a special coordinate condition. 
They have shown that the reduced Hamiltonian describes no 
dynamics within their special choice of coordinate 
condition. 
This conclusion is compared to Birkhoff theorem  in spherically 
symmetric geometry \cite{Weinberg72}. 
The  Wheeler-DeWitt (WD) equation is obtained 
\cite{Wheeler68,DeWitt67} 
 by taking Schr$\ddot{{\rm o}}$dinger picture and by imposing constraints 
  on the state vector.  
The Schr$\ddot{{\rm o}}$dinger picture is adequate to take 
into account of the space-time structure of the universe, 
though it does not incorporate with the unification of elementary 
forces, which is usually described in the Fock space formalism. 

There are serious  problems of the quantum treatment, 
 the definition of the time \cite{Isham92},  
the interpretation of the wave function which is 
a solution of the WD equation\cite{Kuchar93}, 
and the realization of classical system from quantum 
system\cite{Halliwell87}. 


Since it is difficult to treat the WD equation in general, 
the spherically symmetric geometry is often considered. This  
geometry can describe the local 
black hole structure as well as the global cosmological structure, 
though it is not possible to include gravitational waves. 
The Dirac quantization formalism and the WKB approximation 
has been applied to spherically symmetric gravity by 
 Fischler, Morgan, and Polchinski \cite{Fischler90}.  
They derived the WKB wave functions, 
and applied them to the nucleation of matter in the epoch of 
the inflation universe.  

The ADM formalism for the geometro-dynamics of the Schwarzschild 
black holes has been investigated extensively by Kucha$\check{r}$
\cite{Kuchar94}. 
The reduced Hamiltonian was derived as the surface term imposing 
that the constraints held. 
The reduced Hamiltonian method has been developed for the 
Reissner-Nordstr$\ddot{{\rm o}}$m-(anti-)de Sitter 
black holes by Louko et al., and Makela et al.\cite{Louko96,Makela98}. 
The Ashteckar's canonical formalism was also developed for  
the spherically symmetric gravity \cite{Kastrup94,Thiemann95}.
In Ref.\cite{Kastrup94}, an exact Hamilton-Jacobi solution of the 
(classical) WD equation was given.       


 In solving the WD equation, there are operator ordering ambiguities 
 and the regularization problem in defining the product of operators 
 at the same space time point. DeWitt made the regularization scheme that 
 the operator can commute at the same space-time point requiring 
 $\delta (0)=\delta(0)' =0$ \cite{DeWitt67}. 
Schwinger proposed a resolution of the operator ordering problem 
by using some symmetric ordering and  
taking the relation $\delta(0)' =0$ \cite{Schwinger63}. 
 The operator ordering should be taken into account, but 
 usually it is not taken seriously. If the ordering is neglected, 
 the solution of the WD equations valid only  
 in the semi-classical (or WKB) approximation. 
The relation between the operator ordering  and the anomaly 
which arises from 
the algebra among constraints are discussed by Kamenshchik and 
Lyakhovich  in the analogy with string theory \cite{Kamenshchik97}. 
They introduced  a new basis of gravity constraints 
and studied them by using the harmonic expansion 
for cosmological models. They 
got a relation between the space dimension and 
the spectrum of matter fields, just like the string theory. 
Recently Hori gave the solution of the Wheeler-DeWitt equation 
for the Schwarzschild-de Sitter geometry 
under a regularization that any operator commutes 
at the same space-time point, 
and  considered the problem of time by using the Heisenberg 
equation of motion \cite{Hori98}.

In this paper, we consider the black hole solution in spherically 
symmetric geometry. We define the operator ordering for the 
Hamiltonian, the momentum and the mass operators such that 
they satisfy a closed algebra, and derive analytic solutions 
which satisfy all constraints.    

This paper is organized as follows.
In section 2, we review and summarize the canonical formalism 
of the Einstein-Maxwell theory with a cosmological constant 
in spherically symmetric space-time. The derivation of the mass 
function is also described. 
In section 3, we perform the canonical quantization in the 
Schrodinger picture  and obtain the set of constrained wave equations: 
 the momentum constraint equation, the WD equation, and the mass and 
the charge eigen state equations. They are solved consistently after 
fixing the operator orderings.  Discussion is given in section 4.      

%
\newsection{Canonical formalism 
of the Einstein-Maxwell theory  
in spherically symmetric space-time }
%
In this section, we  summarize the canonical formalism 
of the Einstein-Maxwell theory with a cosmological constant 
in the spherically symmetric space-time in four dimensions. 
We use the natural units $c=\hbar=G=1$ and follow mainly 
the convention adopted in Kucha$\check{r}$'s work[13].

We start to consider the general spherically symmetric space-time metric 
in the ADM decomposition   
\bea
    ds^2 = -N^2 dt^2 +\L^2(dr + N^{r} dt)^2 + R^2 d\Omega^2,
    \label{ds^2}
\eea
where $d\Omega^2$ is the line element on the unit sphere, 
the lapse function $N$, the shift vector $N^{r}$, and 
metrics $\L$ and $R$, which  are the function of 
time coordinate $t$ and radial coordinate $r$. 

The action of the Einstein-Maxwell theory is of the form:
\bea
    I = \f{1}{16\pi} \int d^4x \sqrt{-^{(4)}g}
        (^{(4)}R - 2\l -F_{\mu\nu}^2) \hspace{0.5cm} ,
                                             \label{I}
\eea 
where $\l$ denotes the cosmological term, 
and $^{(4)}R $ and $^{(4)}g $ are the scalar curvature and  
the determinant of metrics in four dimensions. 
The electromagnetic field strength is denoted by $F_{\mu\nu}$, which 
is the function of only $t$  and $r$. 
The only non vanishing component is 
\bea
     F_{01} = \dot{A_1} - {A_0}',
     \label{F}
\eea 
where $A_0$ and $A_1$ denote the electromagnetic fields. 
Subscripts, dot and prime 
denote the derivative with respect to $t$ and $r$, respectively. 
As the gravitational field and the electromagnetic field depend 
only on the time and radial coordinates,  this action 
 describes the space-time structure of the universe, but 
does not include the gravitational waves and the electromagnetic 
waves.

Inserting the metrics (\ref{ds^2}) into the action (\ref{I}), 
the action in the ADM decomposition is given as
\bea
    I = \int dt \int dr [
             -N^{-1} (R(-\dot \L + (\L {N^{r}} )') (-\dot R + R' N^{r})
              + \frac{1}{2} \L (-\dot R + R' N^{r})^2 ) 
              \nonumber \\ 
              + N (-\L^{-1}RR'' -\frac{1}{2}\L^{-1}  R'^2 + \L^{-2} R R' \L')
              + \frac{1}{2} N \L (1-\l R^2)  
              \nonumber \\
              + \frac{1}{2} N^{-1} \L ^{-1} R^2  (\dot{A_1} - A_0')^2                             ] \ .
              \label{I-ADM} 
\eea 
The canonical momenta are obtained from the action (\ref{I-ADM}) 
by differentiating with respect to $\dot R, \dot \L, $ and $\dot A_1$ :
\bea 
    P_{\L} &=& - N^{-1} R (\dot R - R' N^{r}) \ \ ,  \nonumber \\
    P_R    &=& - N^{-1} [R (\dot \L - (\L {N^{r}} )') 
                    + \L (\dot R - R' N^{r}) ] \ \ , \nonumber \\
    P_A    &=& N^{-1} \L^{-1} R^2 (\dot{A_1} - A_0') \ \ .
    \label{momenta}
\eea
The fact that the canonical momenta conjugate to 
$N, N^{r}$ and $A_0$ vanish  gives the primary constraints. 
The action can be written in the canonical form after the Legendre 
transformation 
\bea 
    I = \int dt \int d r 
        [ P_R \dot R + P_{\L} \dot \L + P_A \dot A_1 
        - (N H + N^{r} H_{r} + A_0 H_A) ] \ \ ,
       \label{I-canonical}
\eea
where 
\bea
    H      &=& -R^{-1} P_{\L} P_R + \frac{1}{2}R^{-2}\L P_{\L}^2
               + \L^{-1} R R'' - \L^{-2}R R' \L' \nonumber\\
           &&\quad\; +\frac 12 \L^{-1}R'^2-\frac 12 \L
           +\frac{\lambda}2\L R^2+\frac 12\L R^{-2} P_A^2 \ \ , 
           \nonumber \\
    H_{r} &=& R' P_R - \L P_{\L}' \ \ , \nonumber \\
    H_A    &=&  -P_A' \ \ .
    \label{super-Hamiltonian}   
\eea
The quantities $H$ and $H_r$ are called the super-Hamiltonian 
and the super-momentum of gravity. They should vanish 
\bea
    H \approx 0 \ \ , \ \ H_{r} \approx 0 \ \ , 
    \ \ H_A \approx 0 \ .
    \label{constraint}
\eea  
The constraint for $H_A$ corresponds to the Gauss law of 
the electromagnetic field.
 
 The Poisson brackets at the same time $t$ between these constraints 
form a involution relations 
\bea
    \{ H_r (r), \ H_r (r') \} &=& H_{r}(r) \ \d'(r-r') 
    -(r \leftrightarrow r') \ \ ,  \nonumber \\
    \{ H(r), \ H_r (r') \} &=&  H'(r) \ \d(r-r') 
    + H(r) \ \d'(r - r') \ \ ,  \nonumber \\      
    \{ H(r), \ H(r') \} &=& \L^{-2}(r)H_{r}(r) \ \d'(r-r') 
    -(r \leftrightarrow r') \ \ ,     
    \label{PoissonHH} 
\eea 
where $r$ and $r'$ are different space coordinate 
 and $(r \leftrightarrow r')$  
indicates the interchange the argument in the preceding term.
The Poisson brackets between $H_A$ and other constraints 
are trivially vanish. 
These constraints form the first class, that is, 
a kind of gauge theory,
which reflects the general covariance of the spherically symmetric 
space-time.

\vspace{0.5cm}
In order to study the canonical theory of black hole, 
we have  to introduce the mass function. 
The mass function is defined by 
the integration with respect to the space coordinate $r$ 
of a linear combination of the super-Hamiltonian and the super-momentum 
as 
\bea
  &&-\int^{r} dr  
        ( \L^{-1} R' H +  R^{-1} P_{\L}\L^{-1} H_{r} ) \ ,
      \nonumber \\ 
  &&\quad = -\frac{1}{2}( - R^{-1} P_{\L}^2 + \L^{-2} R R'^2 
                    -R-R^{-1}P_A^2 + \frac{\lambda}{3}R^3) -m
                    \nonumber\\
                 && \quad   \equiv M-m \ ,
      \label{M}
\eea 
where $m$ is a integration constant, which is assigned a mass of 
black hole as we see later.       
The physical meaning of the mass function is studied 
by Nambu and Sasaki \cite{Nambu88} identifying with the energy of the inner part of the Schwarzschild black hole.  
The canonical treatment of the mass function was considered 
by Fischler, Morgan, and Polchinski \cite{Fischler90}, and
by Kucha$\check{r}$\cite{Kuchar94} for the Schwarzschild black holes.
The mass function can be considered as a constraint, 
of which Poisson brackets with other constraints at the same 
time $t$ are
\bea
    \{M(r),H_r (r')\}&=& M'(r) \d (r-r') \ , \label{PoissonM} 
    \nonumber\\
    \{M(r),H(r')\}&=& - \L(r)^{-3} R'(r)H_r(r)\d (r-r') \ , 
    \nonumber\\
    \{M(r),M(r')\}&=&0 \ .
    \label{PoissonMM}
\eea
The Poisson bracket shows that the mass function is 
a constant of motion in weak sense.

\newsection{Quantization, operator ordering and analytic solutions of 
Wheeler-DeWitt equation}
In this section we proceed the canonical quantization of 
the  Einstein-Maxwell theory with the cosmological term in the 
spherically symmetric space-time. 

The momenta  corresponding to 
the electromagnetic field $A_1$ and the metrics $\L , R $ 
(\ref{momenta})
are represented by functional differential operators 
in the Schr$\ddot{{\rm o}}$dinger picture 
\bea
    \hat{P}_A    (r) &=& -i \frac{\d}{\d A_1 (r)} \ , \nonumber \\
    \hat{P}_{\L} (r) &=& -i \frac{\d}{\d \L  (r)} \ , \nonumber \\
    \hat{P}_R    (r) &=& -i \frac{\d}{\d R   (r)} \ .
    \label{S-representation}
\eea
In the following we use notation "hat" for the quantized 
differential operators 
and we neglect to express the time $t$ explicitly for 
operators because we treat them at the same time. 
The fact that momenta conjugate to $A_0$ and $N, N^{r}$ 
and their time derivatives  vanish   
leads  the constraint equations (\ref{constraint}). 
In quantum theory, these constrained equations are understood 
as the equations operating on the state vector
\bea
   \hat{H}\Psi=0 \ ,      
    \ \hat{H}_{r}\Psi=0 \ , 
    \ \hat{H}_A\Psi=0  ,     \label{wave equation} 
\eea
and  
\bea
    \hat{M}\Psi = m \Psi \ ,
    \label{mass equation}
\eea 
where the integration constant $m$ plays a role of 
 the eigen value of the mass operator.
The first equations in Eq.(\ref{wave equation}) 
is famous Wheeler-DeWitt (WD) equation for spherically symmetric case. 

It is quite hard to obtain solutions of the above four constraints.  
Furthermore, we have to specify the operator ordering for $\hat H$, 
$\hat H_r$  and $\hat M$ to make the problem well defined. 
We utilize this operator ordering and choose it so that we 
can solve the above four constraint equations. Although we have  
a freedom to specify the operator ordering, it is a quite difficult 
problem to give a consistent ordering in a sense that 
$\hat H$, $\hat H_r$  and $\hat M$ form a closed algebra. 

Our strategy to obtain analytic solutions is as follows:

\begin{description}
\item[(1)] We solve the charge constraint $\hat H_A \Psi=0$ and 
obtain general solutions.
\item[(2)] We specify the operator ordering for $\hat H_r$ and 
then obtain general solutions of the momentum constraint 
$\hat H_r \Psi=0$. 
\item[(3)] We define a specific operator ordering for $\hat M$ such 
that $\hat H_r$ and $\hat M$ form a closed algebra. Then, we 
require the mass constraint $(\hat M -m)\Psi=0$ for 
solutions of momentum constraint. 
\item[(4)] We define the Hamiltonian through the spacial derivative 
of the mass operator, $M'$, which is defined to reduce to the 
classical one in Eq.(2.10) when the ordering is neglected. 
We derive an explicit form of $\hat H$ and 
 show that solutions for three constraints  
$H_A \Psi=0$, $\hat H_r \Psi=0$ and $(\hat M -m)\Psi=0$ 
satisfy automatically the Hamiltonian constraint $\hat H \Psi=0$. 
\item[(5)] We  show that the operators $\hat H$, $\hat H_r$ and $\hat M$ 
form a closed algebra.
\item[(6)] We take a specific operator ordering for $\hat M$ 
and then derive analytic solutions for all four constraints. 
\end{description}

Following the above strategy, we shall explain the procedure 
to solve the problem in detail.

\vskip 5mm
\noindent
(a) General solutions for the charge constraint

The wave function $\Psi$ is a functional of 
the electromagnetic field $A_1$ and the metrics $\L , R $ , 
and is assumed to be in the 
form of separation of variables as
\bea
   \Psi = \Psi_A[A_1] \Psi_G[\L,R] \ ,
   \label{Psi}
\eea
where $\Psi_A$ and $\Psi_G$ denote the wave function of 
the electromagnetic part and that of gravity part respectively.
The Gauss law equation $\hat{H}_A\Psi_A=0$ corresponds to the 
charge conservation and its eigen value equation 
is solved as
\bea
    \hat{P}_A\Psi_A &=& Q \Psi_A \ , \nonumber \\
    \Psi_A[A] &=& \exp (i \int dr \ QA_1(r)) \ ,
    \label{PsiA} 
\eea   
where $Q$ is the eigen value of the conserved charge.


\vspace{0.5cm}

\noindent
(b)  General solutions of the momentum constraint

We defined the operator ordering of $\hat H_r$ 

\bea
    \hat{H}_r  = R' \hat{P}_R -\L (\hat{P}_{\L})'\ ,
\eea
and the momentum constraint equation is explicitly written by
\bea
    \hat{H}_{r}\Psi_G = -i
    (R'(r)\frac{\d}{\d R(r)} - \L (r)(\frac{\d}{\d \L (r)})')
    \Psi_G = 0 \ .
\eea
Note that other form of operator ordering is related to that in Eq.(3.6) 
by the similarity transformation \cite{ordering}.

We define a quantity  $Z$ which is a functional of $R$ and $\L$ 
and is independent of the coordinate $r$, and require that it  
satisfies the momentum constraint equation 
\bea
 [Z, H_r(r)]=i\left( R'(r)\frac{\d Z}{\d R(r)} 
 - \L (r)(\frac{\d Z}{\d \L (r)})'\right )=0\ .
\eea
The solution is expressed by 
\bea
    Z=\int dr  \L f(R,\L^{-1}R')=\int dr 
    \int^{\L}d\L \bar f(R,\L^{-1}R') ,
    \label{Z}
\eea
where $f$ and $\bar f$ is the arbitrary function of $R$ and $\L^{-1}R'$, 
and the integration of $r$ extends between boundaries. 
 The function $f$ and $\bar f$ is related by 
\bea
f(R, \L^{-1} R')=- \L^{-1} R' \int^{R'/\L}
\frac {dx}{x^2} \bar f(R,x)\;.
\eea 
By using $Z$, the solution of the momentum constraint is given 
by 
\bea
    \Psi_G = \Psi_G (Z) \ ,  
    \label{PsiG}
\eea
This solution is a general solution in a sense that 
the functional form of  $f$ or $\bar f$ and also $\Psi_G$ is not 
specified yet. These forms are determined by imposing the 
mass constraint. We comment that 
the most general solution is expressed by the product as
$\Psi_G =\Psi_G(Z_1,Z_2,...)$ where 
$Z_j = \int dr \L f_j(R, \L^{-1}R'), \ (j=1,2,...)$.

\vspace{0.5cm}

\noindent
(c) General solutions of the mass constraint

We defined the operator ordering of $\hat M$ as 
\bea
\hat M-m=\frac 1{2}R^{-1} \hat P_\L^{(A)} \hat P_\L -\frac{1}2 R
(\chi -F)\;  , 
\eea
where  
\bea
\hat P_\L^{(A)}&=& A\hat P_\L A^{-1}\;,\nonumber\\
\chi &\equiv& \L^{-2} R'^2 \;,\;\;
F=1-2mR^{-1}+Q^2 R^{-2}
-\frac{\lambda}{3}R^2\;.
\eea 
The mass operator in Eq.(3.12) reduces to Eq.(2.10) 
when the operator ordering is ignored. The ordering factor 
 $A$ is taken as 
\bea
A=A_Z(Z)\bar A(R,\chi)\;.
\eea

The following comments are in order: For $A$,  we assume 
that it is a function of $Z$, $R$ and $\chi =(\L^{-1}R')^2$ 
in Eq.(3.13). 
The reason of this choice is 
due to the fact that  these operators as well as $\hat P_\L$ are 
good operators in a 
sence that they satisfy the following commutation relation 
with $\hat H_r$
\bea
[\phi(r), \hat H_r(r')]=i\phi'(r)\delta (r-r')\;,
\eea
where $\phi=R$, $R'/\L$, $\hat P_\L$ and $Z$ ($Z'=0$) and a function 
of these quantities. This is needed for $\hat H_r$ and $\hat M$ 
form a closed algebra as we shall see later. 
Then, we assume the factorization between 
a function of $Z$ and a function of $R$ and $\chi$,  which is 
needed to guarantee 
\bea
[\L (r), \hat P_{\L}^{(A)}(r')]=i \delta (r-r')\;,\;\; 
[\hat P_{\L}^{(A)}(r), \hat P_{\L}^{(A)}(r')]= 0\; . 
\eea

The mass operator $\hat M -m$ is applied to solutions in Eq.(3.11) 
of the momentum constraint. 
The mass constraint equation is now given by
\bea
\left(\frac{\delta Z}{\delta \L}\right)^2 \frac{d^2\Psi_G(Z)}{d Z^2}
+A\left[\frac{\delta }{\delta \L}\left(A^{-1}\frac{\delta Z}
{\delta \L}\right)  \right] \frac{d \Psi_G(Z)}{d Z}
+(R\sqrt{\chi-F})^2\Psi_G(Z) =0\;.
\eea  
Now we fix the functional form of $\bar A$  defined 
in Eq.(3.14) and $\bar f\equiv \delta Z/\delta \L$ defined 
in Eq.(3.9) such that the constraint equation in Eq.(3.17) 
does not contain $R$ and $\L$ explicitly, that is, it 
becomes an ordinary differential equation of $Z$. We 
take
\bea
\bar A=\frac{\delta Z}{\delta \L}=
R\sqrt{\chi-F}\;.
\eea
Since  $\bar f$ is fixed, the variable $Z$ is determined. 
Now we only need to determine the functional form of 
$\Psi_G$.  The function $\Psi_G$ satisfies 
\bea
 \frac{d^2\Psi_G}{d Z^2}
-A_Z^{-1}\frac{\delta A_Z}{\delta \L} \frac{d \Psi_G}{d Z}
+\Psi_G =0\;.
\eea
Only the unfixed part is  $A_Z$. We take the form of $A_Z$ 
so that the equation can be solved analytically in 
below.  
\vskip 5mm
\noindent
(d) Hamiltonian

The operator ordering of Hamiltonian is defined through the relation
\bea
(\hat M)'=-\L^{-1}R'\hat H- R^{-1}\hat P_\L^{(B)}
\L^{-1}\hat H_r\;,
\eea
where this relation is a quantum version of Eq.(2.10) and 
\bea
\hat P_\L^{(B)}\equiv \frac 12 
\left (\hat P_\L+\hat P_\L^{(A)}\right )
=A^{1/2}\hat P_\L A^{-1/2}\;.
\eea
 The Hamiltonian is explicitly written by 
\bea
\hat H=\frac{1}{2}\L R^{-2}\hat P_\L^{(C)}\hat P_\L
-R^{-1}\hat P_R\L
\hat P_\L^{(B)}
\L^{-1}+\L{R'}^{-1}\left(\frac{R}{2}(\chi-F)\right)'\;,
\eea
where 
\bea
\hat P_\L^{(C)}\equiv \hat P_\L^{(A)}-iR{R'}^{-1}\left (A^{-1}
\frac{\delta A}{\delta \L}\right)'=C\hat P_\L C^{-1}\;,
\eea
with
\bea
C=A \exp \left[-\int^r dr RR'^{-1}\int^\L d\L\left (A^{-1}
\frac{\delta A}{\delta \L}\right)'\right ]\;.
\eea
In the limit of neglection opereator 
ordering, this Hamiltonian reduces to the classical one 
defined in Eq.(2.7). 

From Eq.(3.20), we see that solutions of $\hat H_r\Psi_{G}=0$ and 
$(\hat M-m)\Psi_{G}=0$ satisfy automatically the Hamiltonian 
constraint $\hat H \Psi_{G}=0$. 

\vskip 5mm
\noindent
(e) Commutation relations

For $\hat H_r$ and $\hat M$ defined in Eqs.(3.6) and (3.12), we find 
\bea
[\hat M(r), \hat M(r')]=0\;, \;\;
[\hat M(r), {\hat H}_r(r')]=i\hat M'(r)\delta (r-r')\;,
\nonumber
\eea
\bea
[{\hat H}_r(r), {\hat H}_r(r')]=i\left(\hat 
H_r(r)\frac{\partial}{\partial r}
\delta (r-r')-\hat H_r(r')\frac{\partial}{\partial r'}
\delta (r-r')\right)\;.
\eea
We comment that the mass operator $\hat M$ in (3.12) consists of 
good operators in a sense that we stated in Eq.(3.15), so that 
the commutator between $\hat H_r$ and $\hat M$ satisfies the 
above simple and closed algebra as we expect form the 
Poisson brackets in Eq.(2.11). 

The commutation relations involving $\hat H$ is complicated, 
but we can derive them by differenciating the above 
commutators among $\hat M$ and $\hat H_r$:
\bea
0&=&\frac{\partial}{\partial r}[\hat M(r), \hat M(r')]\nonumber\\
&=&[-\L^{-1}R'\hat H- R^{-1}\hat P_\L^{(B)}\L^{-1}\hat H_r
, \hat M(r')]\;.
\eea 
We find 
\bea
[\hat H(r),\hat M(r')]&=&i\L^{-3}R'\hat H_r(r)\delta (r-r')
+\L^{-2}R^{-1}\hat H(r)\delta (0)\delta (r-r')
\nonumber\\
&& -\frac 14\frac{\L(r)}{R'(r)R(r)R(r')}
\left\{g(r,r')\hat P_\L(r')-\hat P_\L^{(A)}(r')g(r,r')\right\}
\L^{-1}(r)\hat H_r(r) \;,\nonumber\\
\eea
where
\bea
g(r,r')&=&\frac{\delta}{\delta \L(r)}\left(
A^{-1}(r)\frac{\delta A(r)}{\delta \L(r')}\right)\nonumber\\
&=&\left (A_Z^{-1}A_Z'\right )'\bar f(r)\bar f(r')+
\left\{A_Z^{-1}A_Z'\frac{\partial \bar f}{\partial \L}+
\frac{\partial }{\partial \L}\left(\bar A^{-1}
\frac{\partial \bar A}{\partial \L}\right)\delta (0) \right\}
\delta (r-r')\nonumber\\
&=&g(r',r)\;.
\eea
Here, the dash for $A_Z$ denotes the derivative 
with respect to $Z$ as $A_Z'=d A_Z/dZ$. 
As we expected, the main term $i\L^{-3}R'\hat H_r(r)\delta (r-r')$ 
coincides 
to the one appeared in the Poisson bracket (2.11). Other terms 
that are proportional to $\hat H$ and $\hat H_r$ are those 
due to the operator orderings. Since the commutaror 
between $\hat H$ and  $\hat M$ 
contains only terms proportional to 
$\hat H$ and $\hat H_r$, the mass operator 
$\hat M$ is the conserved quantity. 

By differentiation the commutator between $\hat M(r)$ and 
$\hat H_r(r')$ with respect to $r$ and by using 
the commutator between $\hat H_r(r)$ and 
$\hat H_r(r')$, we find 
\bea
[\hat H(r), \hat H_r(r')]=i\hat H'(r)\delta (r-r')
+i\hat H(r)\delta' (r-r')\;,  
\eea
where $\delta' (r-r')=\partial \delta (r-r')/\partial r$. 
The above commutator is just the quantum extensiton of 
the Poisson bracket in Eq.(2.9). 

By differentiation the commutator between $\hat M(r)$ and 
$\hat M(r')$ with respect to $r$ and $r'$ and by using 
the commutation relations obtained above, we find 
\bea
[\hat H(r), \hat H(r')]=i\L^{-2}(r)\hat H_r(r)\delta' (r-r')
-h(r,r')H_r(r') -(r \leftrightarrow r')\;,
\eea
where
\bea
h(r,r')=\left(\frac{\L}{R'} \right)_{r'}\left[\left(\frac{\L}{2R^2}
\hat P_\L^{(C)}\hat P_\L -R^{-1} P_R
\L\hat P_\L^{(B)}\L^{-1}\right)_r, \left 
(R^{-1}\hat P_\L^{(B)}\L^{-1} \right)_{r'}\right]\;.
\eea
The function $h(r,r')$ is determined as a function of $A_Z$ and 
$\bar A$. 
We do not show the functional form of  $h(r,r')$ explicitly 
here, because it is complicated. The first term in the 
right-hand side is what we expected from the expression of 
the Poisson bracket. The second term is due to our choice of the 
operator ordering. 

We showed all commutators and saw that the algebra is 
closed. The commutators correspond to the Poisson brackets 
in Eq.(2.9), except for the appearance of the terms which are 
proportional to $\hat H_r$. They contain  
$g(r,r')$ and $h(r,r')$ which 
are determined when $A_Z$ and $\bar A$ are given, and 
are the non-local terms in the sense that 
they are not proportional to the delta function and its 
derivatives, unless we take $A_Z^{-1}A_Z'$ to be a constant, 
i.e., $A_Z=\exp \alpha Z$ with $\alpha$ a constant. 
In below, we take $A_Z$ some polynomial of  $Z$, so that 
these non-local terms exist in the commutators.

\vskip 5mm
\noindent
(f) Analytic solutions

We choose $A_Z$ such that the solution of the equation (3.19) 
becomes a special function. 

\vskip 2mm
\noindent
(f-1) Hypergeometric type solutions

If we choose $A_Z=Z^{\sigma}(Z-1)^{\delta}$ and transform
\bea
\Psi_G^{(\sigma,\delta)}(Z)=Z^{\sigma +1}(Z-1)^{\delta +1}\psi(Z)\;,
\eea
we find 
\bea
\psi''+\left(\frac{\sigma+2}{Z}+\frac{\delta+2}{Z-1}\right)\psi'
+\frac{\sigma+\delta+2}{Z(Z-1)}\psi=0\;.
\eea
so that solutions are given by hypergeometric function as
\bea
\Psi_G^{(\sigma,\delta)}(Z)&=&Z^{\sigma +1}(Z-1)^{\delta +1}
\left(a_1F(\alpha,\beta,\gamma;Z)\right.\nonumber\\
&&\quad\left. +a_2Z^{1-\gamma}F(\alpha-\gamma+1,
\beta-\gamma+1,2-\gamma;Z)
 \right)\;,
\eea 
where $a_1$ and $a_2$ are integration constants, and $\alpha$, 
$\beta$ and $\gamma$ are given by
\bea
\alpha\beta=\sigma+\delta+2\;,\;\;\alpha+\beta=\sigma+\delta+3\;,
\;\;\gamma=\sigma+2\;.
\eea 

\vskip 5mm
\noindent 
(f-2) Bessel type solutions

As a simpler case, we take 
$A_Z=Z^{2\nu-1}$. Then,  the equation becomes
\bea
 \frac{d^2\Psi_G(Z)}{d Z^2}
-\frac{2\nu-1}{Z} \frac{d \Psi_G(Z)}{d Z}+\Psi_G(Z) =0\;.
\eea
Solutions of Eq.(3.32) are given by 
 the Hankel (or Bessel) functions as 
\bea
    \Psi_G^{(\nu)}(Z) = Z^{\n} \ 
                (b_1 \ H_{\n}^{(1)}(Z)+b_2 \ H_{\n}^{(2)}(Z))
        \ ,
        \label{exact wave function}
\eea
where  $b_1$ , $b_2$ are integration constants.   
In a special case of   $\n =1/2$, solutions are  simply expressed as 
\bea
    \Psi_G^{(1/2)}(Z) = \sqrt{\frac{2}{\pi}} \ 
                (b_1 \ \exp{(i(Z-\pi/2))} + b_2 \ \exp{(-i(Z-\pi/2))})\ .
\eea

\vspace{0.5cm}
%
\newsection{Discussion}
%
We have studied the canonical quantum theory of the Einstein-Maxwell 
model with a cosmological term in spherically symmetric space-time. 
We defined the operator orderings for $\hat H$, $\hat H_r$ and
$\hat M$ and showed that they form a closed algebra. 
The commutators reproduced all terms appeared in the 
Poisson bracket. In addition, we found 
some terms which are proportional to $H_r$. These  terms 
contain in general the non-local terms in a sense that 
they are not proportional to $\delta$ 
function and its derivatives.  In this operator ordering, we  
derived analytic solutions which satisfy all constraints, 
the Hamiltonian, the momentum and the mass constraints. 
In our knowledge, this work is  the first to obtain 
solutions of these constraints in the spherically symmetric 
geometry, with the operator ordering such that the algebra 
is closed. 

From the point of the quatnum field theory, the field operator 
in the  the Schr$\ddot{{\rm o}}$dinger picture is generally 
difficult to define consistently and mathematically. 
For general relativity, it is difficult to solve 
the Wheeler-DeWitt equation even formally. 
Our procedure succeeded in bypassing these difficulties 
through the mass operator, which is the integral of the 
super-Hamiltonian and the super-momentum. 

We now compare our solutions and the WKB (semi-classical) solution. 
  The WKB wave function is given by\cite{Fischler90,Kastrup94,Hori98} 
\bea
    \Psi_G \sim c_1 \exp{ (iZ_{\rm WKB})}
             +c_2 \exp{(-iZ_{\rm WKB})} \ ,\nonumber
\eea
where $c_1$, $c_2$ are integration constants and 
\bea
     Z_{\rm WKB}&=& 
       \int d r \int ^{\L} d\L \ R \sqrt{\chi - F} 
       \ ,  \nonumber\\
        &=& \int dr R\L ( \sqrt{\chi - F} + 
         \sqrt{\chi}  \log \mid 
          \frac{\sqrt{-F}}{\sqrt{\chi} 
          +\sqrt{\chi - F}} \mid ) 
           \ , \ \ ({\rm for} \ F<0) \ , \nonumber
\eea    
where $\chi$ and $F$ are defined in Eq.(3.13). The function  
$Z_{\rm WKB}$ turns out to be the same as our $Z$ in Eq.(3.9) and 
the WKB solution agrees with our Bessel type solution for $\nu=1/2$.  
Our Bessel type solutions 
with arbitrary $\nu$ approach to the WKB solution 
for large values of $Z$. 

Our solutions are rather general since we did not impose 
any coordinate conditions. All  solutions that are known 
before are obtained by imposing some coordinate conditions. 
Thus,  we limit the geometry such as Schwarzschild geometry or 
the de Sitter geometry, and 
impose some coordinate conditions, and see whether  our solutions   
reduce to  some known  solutions. Since our solutions are 
given as a function of $Z$, this limiting procedure is 
the process to find the functional form of $Z$. 

\noindent 
(i) The Schwarzschild black hole case  ($Q=\lambda=0$):

Further, the coordinate conditions, 
 $R'=0$ so that $\chi=\Lambda^{-2}R'^2=0$, and $\dot{R}=1$ 
 are imposed. Then, we require $\Lambda=\sqrt{-F}$ in addition,  
the function $Z$  is determined by 
\bea
Z=\int dr \Lambda R \sqrt{\chi-F} \mid_{\chi=0,\Lambda=\sqrt{-F}}
   = \int dr (-R+2m)\;.\nonumber
\eea
The Bessel type solutions with the above $Z$ agree with the 
ones given by Nakamura et al.[23] and by our previous work[24]. 
This case is extended to the case where $Q, \lambda \neq 0$. 
In our recent work[25],  solutions are given and the interpretation 
 of wave functions  have been given.  

\noindent
(ii) The de Sitter universe case ($m=Q=0$):

The coordinate conditions, $\chi=1$ and
$\dot{R}=\sqrt{\lambda/3}R'$ are imposed further. If we require 
the relation $\Lambda = \sqrt{\lambda/3}R = a(t)e^{r}$, 
where $a(t)= \exp{(\sqrt{\lambda/3}t)}$, 
$Z$ is determined as $Z \propto a(t)^3$.  If we choose 
the index of the Bessel type solution as $\nu=1/3$ (Airy function), 
our  solution  agrees with that of Horiguchi[26]. 

In summary, our solution can describe the local structure such 
as a black hole and the global structure such as universe.

Our solutions may be used to investigate the early universe 
in the inflation time. Also the solutions may be used 
for the analysis of the quantum universe and the quantum 
black hole. In order to study the quantum epoch beyond 
the semi-classical region, we assume some interpretation 
of quantum universe. The de Broglie-Bohm (pilot wave, 
quantum potential) interpretation[27] 
lies in an unique position in a sense that this provide 
the notion of trajectory in the fully quantum region. 
The study of the quantum black hole is now under 
investigation by using the de Broglie-Bohm interpretation.

\clearpage

\clearpage
\pagebreak

\begin{thebibliography}{99}
\bibitem{ordering}
If we take the symmetric ordering for the momentum constraint as
\bea
&&\hat{H}_r^{({\rm sym})}\Psi^{({\rm sym})}=0 \ , \nonumber \\
&&\hat{H}_r^{({\rm sym})} \equiv \frac{1}{2}(R'\hat{P}_R + \hat{P}_R R'
      - \Lambda \hat{P}_{\Lambda} - \hat{P}_{\Lambda} \Lambda)
 \ , \nonumber
\eea
then the operator and the wave function are related to those in 
Eqs.(3.6) and (3.7) by the similarity transformation as 
\bea
&&D^{-1}\Psi^{({\rm sym})} = \Psi \ , \nonumber \\
&&D^{-1}\hat{H}_r^{({\rm sym})} D = \hat{H}_r = 
  R'\hat{P}_R - \Lambda \hat{P}_{\Lambda}
 \ , \nonumber \\
\eea
where 
\bea
D=\exp{[-\frac{1}{2}\int dr \ln R'(r)\Lambda(r)]} \ .
\eea
\end{thebibliography}
\end{document}